\documentclass[aps,prl,twocolumn,showpacs,superscriptaddress,floatfix]{revtex4}
\usepackage{graphicx}
\usepackage{float}

\begin{document}

\title{On the transmission of light through a single rectangular
hole}
\author{F. J. Garc\'{\i}a-Vidal}
\affiliation{\mbox{Departamento de F\'{\i}sica Te\'{o}rica de la
Materia Condensada, Universidad Aut\'onoma de Madrid, E-28049
Madrid, Spain}}
\author{Esteban Moreno}
\affiliation{\mbox{Departamento de F\'{\i}sica Te\'{o}rica de la
Materia Condensada, Universidad Aut\'onoma de Madrid, E-28049
Madrid, Spain}}
\author{J. A. Porto}
\affiliation{\mbox{Departamento de F\'{\i}sica Te\'{o}rica de la
Materia Condensada, Universidad Aut\'onoma de Madrid, E-28049
Madrid, Spain}}
\author{L. Mart\'{\i}n-Moreno}
\affiliation{\mbox{Departamento de F\'{\i}sica de la Materia Condensada,
ICMA-CSIC, Universidad de Zaragoza, E-50009 Zaragoza, Spain}}

\begin{abstract}

In this Letter we show that a single rectangular hole exhibits
transmission resonances that appear near the cutoff wavelength of
the hole waveguide. For light polarized with the electric field
pointing along the short axis, it is shown that the
normalized-to-area transmittance at resonance is proportional to
the ratio between the long and short sides, and to the dielectric
constant inside the hole. Importantly, this resonant transmission
process is accompanied by a huge enhancement of the electric field
at both entrance and exit interfaces of the hole. These findings
open the possibility of using rectangular holes for spectroscopic
purposes or for exploring non-linear effects.
\end{abstract}


\maketitle

% main text

Since the pioneering work of Ebbesen {\it et al.} \cite{Ebbe98}
reporting extraordinary optical transmission (EOT) through
two-dimensional (2D) hole arrays perforated in optically thick
silver films, the study of the transmission properties of
subwavelength apertures (holes or slits) has become a very active
area of research in electromagnetism. In the last few years, EOT
has been found in other frequency regimes, as THz
\cite{Jaime,Grisch} and microwave \cite{beruete}. Nowadays there
is a wide consensus that EOT in hole arrays is linked to the
excitation of the surface electromagnetic (EM) modes
\cite{LMM01,Barnes} that decorate the surfaces of structured
metals.

Very recently, different experiments have focused on the influence
of hole shape on the transmission properties of both 2D hole
arrays \cite{Gordon,Kuipers,Cao} and single subwavelength holes
\cite{Degiron} in the optical regime. All these studies show
strong polarization dependencies in rectangular holes and also
that rectangular holes exhibit higher transmittance than square or
circular holes with the same area. Interestingly, it was also
found that single rectangular holes can support transmission
resonances, even in the subwavelength regime.

In this Letter we present the, up to our knowledge, first
theoretical study on the dependence on hole shape of the
transmittance through a single hole. We find that, as a difference
with circular holes \cite{Roberts}, single rectangular holes can
exhibit strong transmission resonances in all frequency ranges.
One of these resonances appears close to cutoff, with a peak
transmittance controlled simply by the ratio between the long and
short sides of the rectangle. Additionally, we show that the
presence of a material filling the hole greatly boosts the
transmittance. Associated to these transmission resonances, there
is a very strong increase of the electric field at the hole.

\begin{figure}[h]
\begin{center}
\includegraphics[width=\columnwidth]{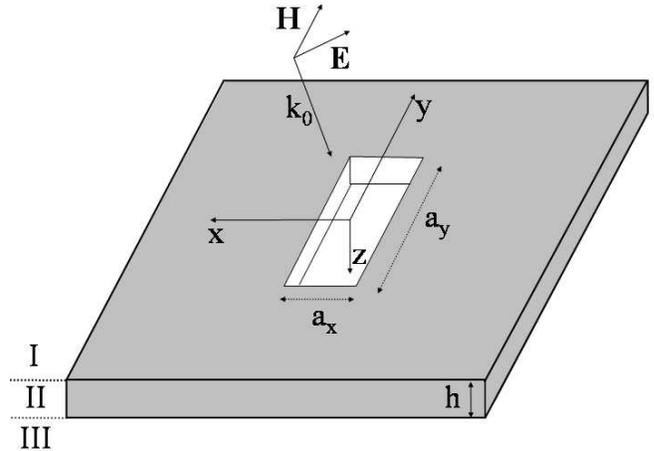}
\end{center}
\caption{Diagram of a single rectangular hole of sides $a_x$ and
$a_y$ perforated on a metal film of thickness $h$. The structure
is illuminated by a $p$-polarized plane wave with the magnetic
field pointing along the y-direction.}
\end{figure}

Figure 1 shows schematically the system under study. A rectangular
hole of sides $a_x$ and $a_y$ perforated on a metallic film of
thickness $h$. The system is illuminated by a $p$-polarized plane
wave with wavelength $\lambda$, the in-plane component of the
electric field pointing along the x-direction. The metal is
treated within the perfect conductor approximation (PCA). This
approximation is known to fail in two instances: first, when the
dimensions of the structure are of the order or smaller than the
skin depth \cite{Sambles04}. Even in this case, the range of
validity of the PCA can be greatly extended by simply considering
effective hole dimensions enlarged by the (metal and wavelength
dependent) skin depth \cite{LMM04SSP}. Secondly, in the optical
regime there are effects related to the existence of absorption
and surface plasmons, which are not captured by the PCA. However,
the PCA serves both as the starting point for more elaborated
approximations, (as the one considering surface impedance boundary
conditions) and for clarifying which effects come from geometry
and which have a dielectric origin. With these caveats in mind,
especially important in the optical regime, the results presented
in this letter apply to different frequency regimes, by simply
re-scaling all lengths by the same factor.

Let us briefly describe the formalism used for calculating the
transmittance through the structure (a detailed account of this
method that was developed to treat an arbitrary number of
indentations can be found in \cite{Bravo}). In this method, the EM
fields in both reflection (I) and transmission (III) vacuum
regions are expressed in terms of EM eigenmodes $|\vec{k}\sigma>$,
characterized by the in-plane component of the wavevector
$\vec{k}$, and the polarization $\sigma$. Inside the hole, the EM
field is expanded in terms of all EM waveguide eigenmodes. After
matching appropriately the EM fields at the two interfaces of the
system ($z=0$ and $z=h$), the formalism provides the full EM field
in all space in terms of the projection onto waveguide eigenmodes
of the electric field at both hole entrance and exit interfaces.
In all calculations presented in this letter (normal incidence
illumination and $a_x, a_y < \lambda$), we have checked that
considering just the first TE eigenmode ($|TE>$) is enough to
obtain very accurate results for the transmittance so, for
simplicity, we present our formalism just for this case. Then, in
terms of the modal amplitudes $E$ and $E^{\prime}$, the electric
field bivector $\vec{E} = (E_x, E_y)^T$ ($T$ standing for
transposition) at the hole entrance and exit can be written as
$|\vec{E}(z=0)> = E \, |TE >$ and $|\vec{E}(z=h)> = - E^{\prime}\,
|TE >$, respectively. Here we have used Dirac's notation, with
$<\vec{r}\,|TE>= (1, 0)^T \sin[\pi(y/a_y+1/2)] / \sqrt{\cal{N}}$,
 ${\cal{N}}= a_x a_y /2$ being a normalization factor.
The equations that $E$ and $E^{\prime}$ must satisfy are:
\begin{eqnarray}
(G-\Sigma)\, E-G_V\, E^{\prime}&=&I_0, \nonumber \\
-G_V\, E+(G-\Sigma)\, E^{\prime}&=&0 \label{system_of_eq}.
\end{eqnarray}

\noindent where $I_0= 2 \imath <\vec{k}_0 p\, |TE>$ is the
external illumination term, i.e.,  the overlap integral between
the incident plane wave $|\vec{k}_0 p>$ and mode $|TE>$. For the
case of normal incidence, and normalizing the incident EM field
such that the incoming energy flux over the hole area is unity, a
simple calculation gives $I_0=4\sqrt{2}\imath/\pi$. $\Sigma$ and
$G_V$ are two magnitudes that only depend on the characteristics
of TE mode inside the hole: $\Sigma=Y_{TE}/\tan(q_z h)$ and
$G_V=Y_{TE}/\sin(q_z h)$; $q_z=\sqrt{k_{\omega}^2-(\pi/a_y)^2}$ is
the propagation constant of the fundamental TE mode, $Y_{TE}=
q_z/k_\omega$ its admittance and $k_\omega=2 \pi/ \lambda$.

The self-illumination of the hole, via vacuum modes, is controlled
by $G= \imath /(2 \pi) ^2 \sum_{\sigma} \int d\vec{k} Y_{\vec{k}
\sigma}|<TE|\vec{k} \sigma >|^2 $ (with $Y_{\vec{k} \sigma}$ being
the admittance of the EM vacuum mode $|\vec{k}\sigma>$
\cite{admit}). For the case of rectangular holes,

\begin{eqnarray}
G&=& \frac{\imath a_x a_y}{8 \pi^2 k_\omega} \! \!
\int_{-\infty}^{+\infty} \! \! \int_{-\infty}^{+\infty} \! \! dk_x
\, dk_y \, \frac{k_{\omega}^2-k_y^2}{\sqrt{k_{\omega}^2-k^2}} \,
{\rm sinc}^2(\frac{k_x a_x}{2}) \nonumber
\\ & &  \,  \,  \, \,  \,  \, \times \,
[{\rm sinc}(\frac{k_y a_y+\pi}{2}) + {\rm sinc}(\frac{k_y
a_y-\pi}{2})]^2
\end{eqnarray}

\noindent where $k^2=k_{x}^2+k_{y}^2$. Notice that, in our
formulation, $\rm{Re}(G)$ comes from the coupling to evanescent
modes in vacuum and $\rm{Im}(G)$ from the radiative modes.

After obtaining $E$ and $E^{\prime}$ from (1), the
normalized-to-area transmittance ($T$) is calculated as:

\begin{equation}
T=\frac{G_V}{Y_{\vec{k}_{0} p}}{\rm Im}[E^*E^{\prime}]
\end{equation}
%\noindent where $Y_{\vec{k}_{0} p}$ is the admittance associated
%with the incident plane wave.

Figures 2 and 3 illustrate the dependence of $T$ with the two
geometrical parameters involved (ratio $a_y/a_x$ and thickness
$h$). In both cases we consider normal incidence. As previously
said, all results are scalable, and we have chosen as unit length
the cutoff wavelength of the TE waveguide inside the hole,
$\lambda_c=2 a_y$.

\begin{figure}[h]
\includegraphics[width=\columnwidth]{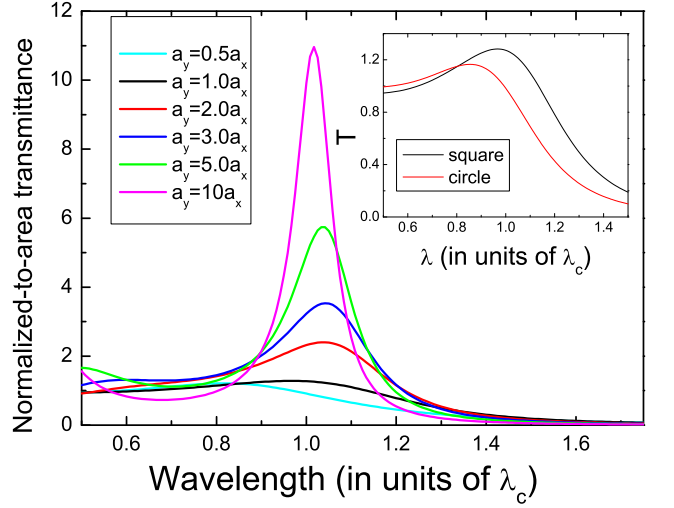}
\caption{Normalized-to-area transmittance (T) versus wavelength
(in units of the cutoff wavelength, $\lambda_c$), for a normal
incident plane wave impinging at a rectangular hole, for different
ratios $a_y/a_x$. Metal thickness is $h=a_y/3$. For comparison, in
the inset we plot $T$ versus wavelength for a single square (black
line) and circular (red line) holes.}
\end{figure}

Figure 2 renders $T$ for the case $h=a_y/3$, for different values
of $a_y/a_x$. As clearly seen in this figure, a transmission peak
develops at approximately $\lambda_c$, with increasing maximum
transmittance and decreasing linewidth as $a_y/a_x$ increases. In
the case of square or circular holes there is also a resonance
close to cutoff, but a very faint one (see inset of Fig.2). In
these last two cases, bellow cutoff the normalized-to-area maximum
transmittance is of order of $1$, i.e., approximately the amount
of light that is directly impinging at the hole. In all cases,
above cut-off $T$ decreases strongly with $\lambda$, both due to
the fact that the fields inside the hole are evanescent and that,
in the extreme subwavelength regime, an incident wave couples very
poorly to the hole \cite{Bethe}.

\begin{figure}[h]
\includegraphics[width=\columnwidth]{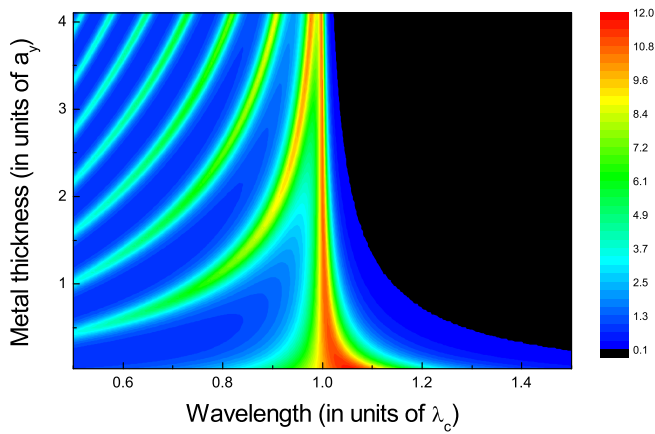}
\caption{$T$ for a normal incident p-polarized plane wave versus
wavelength (in units of $\lambda_c=2a_y$) and thickness $h$
(ranging from $0$ two $4a_y$) for rectangular holes with aspect
ratio $a_y=10a_x$.}
\end{figure}

The dependence of $T(\lambda)$ with metal thickness is shown in Fig. 3,
for a rectangular hole with $a_y/a_x=10$. Apart from the previously
discussed transmission peak located at $\lambda\approx \lambda_c$,
which spectral position essentially remains independent of $h$,
a series of transmission resonances emerge as the depth of the hole is
increased. As we will show later on, these are
Fabry-Perot resonances in
rectangular holes, similar to the ones recently found in
sub-wavelength 1D slits \cite{Porto,Sambles,Bravo2}.

The resonant characteristics of the transmittance through
rectangular holes, and their dependence on geometrical parameters
can be worked out analytically from set of equations (1). For the
case we are analyzing (a symmetric structure with respect to the
plane $z=h/2$), maximum transmission appears when the field
intensities at the entrance and exit sides of the aperture are
equal, i.e. $|E|=|E^{\prime}|$. From (1), this occurs when
$|G-\Sigma|=|G_V|$, condition that after some straightforward
algebra implies

\begin{equation}
2 {\rm Re}(G)=\frac{|G|^2-Y_{TE}^{2}}{Y_{TE}}\tan(q_zh)
\end{equation}

There are several wavelengths at which this transcendent equation
is satisfied, providing the exact spectral location of the
transmission peaks ($\lambda_{res}$) found in Figs. 2 and 3.
Ignoring the shift in the spectral dependence due to the EM
coupling to vacuum modes (this is, setting $G \rightarrow 0$,
which is the appropriate limit for $a_x,a_y << \lambda$), Eq. (4)
transforms into $Y_{TE} \tan(q_zh)=0$, which is the usual
Fabry-Perot condition for the existence of a standing wave inside
the hole. Note that this last equation allows the solution
$q_z=0$, so a transmission peak located at around the cutoff
wavelength is expected, irrespective of the geometrical parameters
$a_y/a_x$ (see Fig.2) and $h$ (see Fig.3). In general, Eq. (4)
predicts the shifts in the Fabry-Perot resonant wavelengths due to
the coupling to both radiative and evanescent vacuum modes.

Using the resonance condition [Eq.(4)], from  Eqs.(1) and (3) and
after some straightforward algebra, we obtain that the
normal-incidence $T$ {\it at resonance}, $T_{res}$ is given by:

\begin{equation}
T_{res}=\frac{|I_0|^2}{4 {\rm Im}(G)}
\end{equation}

A very good simple approximation to $T_{res}$ can be obtained
realizing that, in the extreme subwavelength limit ($a_x, a_y <<
\lambda$), Eq. (2) gives ${\rm Im}(G) \approx 32 a_xa_y/(3\pi
\lambda^2)$. We have checked that this expression holds even for $
a_y \approx \lambda/2$ therefore, for $\lambda > 2 a_y$ we find

\begin{equation}
T_{res} \approx \frac{3}{4\pi}\frac{\lambda_{res}^2}{a_xa_y}
\end{equation}

Recalling that $T$ is the normalized-to-area transmission, this
expression implies that the {\it total} amount of light emerging
from a rectangular hole is, at least for the resonance appearing
close to cutoff, independent of the length of the short side!
Although derived for rectangular holes, Eq. (6) seems to be more
general as the same expression was found for circular holes
\cite{Leviatan}, with the term $a_xa_y$ replaced by the area of
the circular hole. The important point in rectangular holes is
that, for the polarization chosen, the transmittance peak
appearing at cutoff depends only on the long side
($\lambda_{res}=2a_y$), resulting in a transmittance
$T_{res}\approx (3/\pi) a_y/a_x$ close to cutoff. This is the main
result of this Letter, as it predicts a huge transmission
enhancement in a {\it single} rectangular hole with large aspect
ratio.

\begin{figure}[h]
\includegraphics[width=\columnwidth]{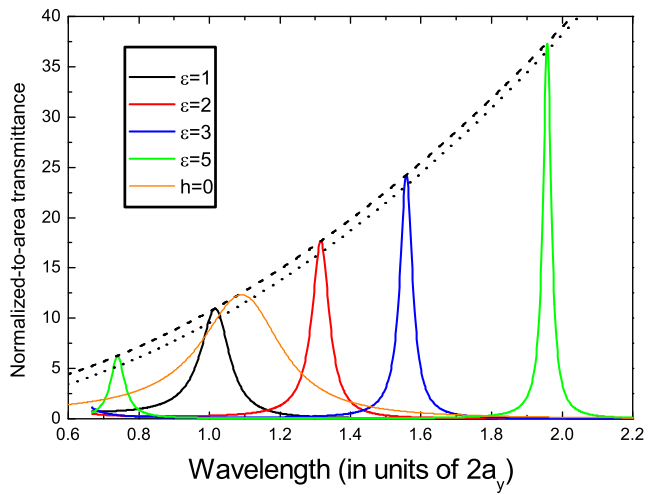}
\caption{$T$ for a normal incident plane wave versus wavelength
for a rectangular hole with $a_y/a_x=10$ and different values of
$\epsilon$ inside the hole. Metal thickness is $h=a_y/3$. Dashed
and dotted lines show the behavior of Eqs.(5) and (6),
respectively. Orange curve shows $T$ versus $\lambda$ in the limit
$h \rightarrow 0$ for a rectangle with aspect ratio $a_y/a_x=10$.}
\end{figure}

Additionally, even for a fixed aspect ratio $a_y/a_x$, Eq. (6)
gives us a clue for further enhancing the transmission: filling
the hole with a material with dielectric constant $\epsilon > 1$,
as this increases the cutoff wavelength. In fact, in the
definition of quantities appearing in Eqs. (1) and (3), the only
place in which $\epsilon$ enters is in the propagation constant
associated to mode $|TE>$ which now reads: $q_z=\sqrt{\epsilon
k_{\omega}^2-(\pi/a_y)^2}$. Therefore, the spectral position of
resonances depend on $q_z$ (and therefore on $\epsilon$) but the
transmittance at resonance is still given by Eq. (5). As a result,
filling the hole with a dielectric would redshift the Fabry-Perot
transmission peak appearing close to cutoff to $\approx 2
\sqrt{\epsilon} a_y$ (for thick enough metal films, see below)
and, more importantly, increase its transmittance. This is
illustrated in Fig. 4, which renders the transmission spectra for
rectangular holes of aspect ratio $a_y/a_x=10$, in a metallic film
of thickness $h=a_y/3$, for several values of $\epsilon$. Note
that this way of increasing the transmission through the hole by
filling it with material with $\epsilon > 1$ can be also operative
for the case of circular \cite{deAbajo} or square holes.
Remarkably, in rectangular holes this mechanism acts almost
independently of the enhancement due to the aspect ratio, so
$T_{res}$ is proportional to both $a_y/a_x$ and $\epsilon$.

It is worth analyzing the limit $h \rightarrow 0$ of the
transmissivity of single rectangular holes, the analogue of
Bethe's study for circular holes \cite{Bethe}. By taking this
limit in the set of equations (1) and Eq.(3), we can obtain an
expression for $T$ valid for all wavelengths:

\begin{equation}
T=\frac{|I_0|^2}{4|G|^2}{\rm Im}(G),
\end{equation}

\noindent which, as expected, is independent of $\epsilon$. The
behavior of this magnitude as a function of wavelength is rendered
in Fig. 4 (orange curve) for the case $a_y=10a_x$. As metal
thickness is decreased, the resonance moves from a location close
to $2\sqrt{\epsilon}a_y$ to a much shorter wavelength, given by
the resonant condition ${\rm Re}(G)=0$ (notice that Eq.(7) at
resonance gives the same expression as Eq.(5) for this case).
Interestingly, in the extreme subwavelength limit, as ${\rm Re}(G)
\propto \lambda$ and ${\rm Im}(G) \propto 1/\lambda^2$, $T$
decreases with $\lambda$ as $1/\lambda^4$, as in the case of
circular holes \cite{Bethe}.

It is also interesting to analyze the enhancement of the EM-fields
associated to this resonant phenomenon. Naively, one would expect
that the intensity of the E-field at the entrance and exit sides
of the hole ($|E|^2$ and $|E^{\prime}|^2$) should be proportional
to the transmittance. However, the direct evaluation of
$|E|^2=|E^{\prime}|^2$ at the resonant condition given by Eq. (4)
yields:

\begin{equation}
|E|^{2}_{res}=|E^{\prime}|_{res}^{2}=\frac{|I_0|^2}{4[{\rm
Im}(G)]^2},
\end{equation}

\noindent leading to an enhancement of the intensity of the
E-field (with respect to the incident one) that scales with
$\lambda_{res}$ as $\lambda_{res}^{4}/(a_y a_x)^2$, square of the
enhancement in the transmittance (see Eq.(6)). This implies that
in the process of resonant transmission, light is highly
concentrated on the entrance and exit sides of the hole but only a
small fraction of this light is finally transmitted.

These effects (huge enhancements of both transmission and field
amplitude) should be readily observable in the microwave or THz
frequency ranges. In these regimes, holes with very large aspect
ratio can be manufactured and, furthermore, dielectric materials
presenting large positive dielectric functions are also available.

We would like to end with some comments about the transferability
of our results to the optical regime. Although our results agree
with the experimental finding that transmittance resonances
increase and become better defined with increased aspect ratio,
they have only a semi-quantitative value in this regime,
especially for short sides not much larger than twice the skin
depth (i.e. $\approx 50nm$). Additionally, the influence of
localized surface plasmon modes on the transmittance is not
capture by our model. The role played by these modes on the
transmittance is a point that deserves further theoretical
investigation.

Financial support by the Spanish MCyT under contracts
MAT2002-01534 and MAT2002-00139, and the EC under Projects No. No.
FP6-2002-IST-1-507879 and FP6-NMP4-CT-2003-505699 is gratefully
acknowledged.

\end{document}